\documentstyle[preprint,aps]{revtex}
\begin{document}
\draft
\preprint{\begin{tabular}{l}
\hbox to\hsize{November, 1996 \hfill SNUTP 96-011}\\[-3mm]
\hbox to\hsize{hep-ph/96xxxxx \hfill Brown-HET-1062}\\[5mm] \end{tabular} }

\bigskip

\title{ Almost maximally broken  permutation 
\\ symmetry for neutrino mass matrix }
\author{Kyungsik Kang}
\address{Department of Physics, Brown University\\ Providence,
Rhode Island 02912, USA}
\author{Sin Kyu Kang and Jihn E. Kim}
\address{Center for Theoretical Physics and Department of
Physics\\ Seoul National University, Seoul 151-742, Korea}
\author{Pyungwon Ko }
\address{Department of Physics, Hong-Ik University \\ Seoul 121-791, Korea}
\maketitle
\tighten
\begin{abstract}
Assuming three light neutrinos are Majorana particles, we propose mass 
matrix ansatz for the charged leptons and Majorana 
neutrinos with family symmetry $S_{3} $ broken into $S_{1}$ and 
$S_{2}$, respectively.    Each matrix has three parameters, which are 
fixed by measured charged lepton masses, differences of squared
neutrino masses relevant to the solar and the atmospheric neutrino 
puzzles, and the masses of three light Majorana
neutrinos as a candidate for hot dark matter with $\sum |m_{\nu}| \sim 6~$eV. 
The resulting neutrino mixing is compatible with 
the data for the current upper limit, 
$ \langle m_{\nu_e} \rangle_{th} < 0.8~{\rm eV}$,
of neutrino-less double beta decay experiments, 
and the current data for various types of neutrino oscillation experiments.
One solution of our model predicts that 
$\nu_{\mu} \rightarrow \nu_{\tau}$ oscillation probability is about
$< 0.008$ with $\Delta m^{2} \sim 10^{-2}~{\rm eV}^2$, which may not be 
accessible at CHORUS and other ongoing experiments.
\end{abstract}


\newpage
\narrowtext
 \tighten

{\bf 1.} 
One of the important problems in particle physics is to understand the
origin of fermion masses and flavor mixing and their hierarchical patterns.
There are several different approaches to this problem, such as  chiral 
flavor symmetries, texture  zeros, mass matrix ansatz, determination of 
the top quark mass using the arguments based on infrared stable 
quasi-fixed point or the multiple point criticality principle, and so on
\cite{raby}. 

Recently,  Fritzsch and his collaborators \cite{nagoya} suggested 
quark mass matrix ansatz based on flavor democracy and its suitable 
breaking.   This idea of flavor democracy is by no means new, and noticed
by several authors before \cite{meshkov}.  
Their work  on the quark sectors \cite{nagoya}
yields the weak  CP phase to be maximal ( $\alpha \approx 90^{\circ}$) 
\cite{ms} in the mass matrix in order for the CKM matrix elements to be 
consistent with the current observations \cite{pdg}.
In view of the undergoing B-factory projects, it is quite interesting to see 
if this prediction of maximal CP-violation is really realized in the 
CP asymmetry in the B decays.  These authors also considered 
the lepton mixing
matrix without flavor democracy in the neutrino sector, and got some
interesting predictions \cite{xing}, which is essentially similar to
the maximal  mixing scenario.

In this letter, we extend Fritzsch's approach to the lepton sector with 
a suitable modification which is different from Ref.~\cite{xing}. 
As discussed below, the flavor democracy predicts one heavy fermion 
versus two massless fermions. This picture nicely fits with the
charged fermion sectors, but not with the neutrino sector.  If one requires
three neutrinos to be a hot dark matter component of the
universe, and solve the solar and the atmospheric neutrino puzzles in terms
of neutrino oscillations, one has to invoke three almost degenerate 
neutrinos, instead of a large hierarchical structure in $m_{\nu_i}$ ($i = 1,
2,3$). Therefore, we relax `the condition of flavor democracy' in the Yukawa 
couplings and require the Yukawa matrices to possess  `the permutation 
symmetry in the family indices.'   Then, the  flavor democratic structure 
turns out to be a special case of the structure with the permutation symmetry.

We first recapitulate the idea by Fritzsch {\it et al.} in the context of 
the charged lepton mass matrix ($M_{l}$).  
Then, we introduce a mass matrix ansatz  for three light neutrinos ($M_{\nu}$) 
invoking the permutation symmetry among three family indices ($S_3$), and its 
breaking to $S_2$ or $S_1$. 
Each lepton matrix depends on three independent real parameters, 
if we assume there is no particular relations among lepton masses.
The form of the lepton mass matrix with three independent real parameters 
may be completely arbitrary in principle, as long as they correctly
reproduce three lepton masses.  In order to reduce this arbitrariness in the
form of the mass matrix ansatz, we assume that the mass matrix ansatz
respect the permutation symmetry among three families ($S_3$) in the 
zeroth order approximation, and then this $S_3$ symmetry is subsequently
broken to a smaller group $S_2$ or $S_1$. This requirement of permutation
symmetry reduces the arbitrariness of our approach based on specific
ansatz for lepton masses.
Thus, there are six real parameters in total in the lepton mass matrices,  
and there is no CP violation in the leptonic sector in our model.
These six parameters are fit to (i) the charged lepton masses, 
(ii) $\Delta m^2$ relevant to the solar and the 
atmospheric neutrino puzzles, and finally (iii) $\sum_{i=1}^{3} | m_{\nu_i} |
\sim 6~$ eV  (in order to solve the dark matter problem) \cite{gelmini}.  
Diagonalizing the mass matrices yields  a neutrino mixing matrix ($V$).  
By studying the resulting predictions to the various 
neutrino  oscillation probabilities, and the neutrino-less double beta decay,
we are led to make an ansatz for the neutrino mass matrix which fits
every known constraint. It also predicts that $P(\nu_{\mu} \rightarrow 
\nu_{\tau}) \sim 0.008$, which is beyond the reach of the CHORUS and 
other similar experiments.

When one considers massive neutrinos, one could think of variety of models
with massive neutrinos including see-saw mechanism and GUT.  In this work,
however, we simply assume that the standard model (both particle contents and 
the symmetry group structure) is valid up to some scale
$\Lambda \ll M_{Z}$. 
Then there is a dimension-5 operator $h_{ij} l_{i} l_{j} H^2 
/ \Lambda$, which generates the Majorana masses for three neutrinos 
$m_{\nu} \sim h v^2 / \Lambda$ after the electroweak symmetry breaking
\cite{ellis}.
If $\Lambda \sim M_{\rm Planck}$,  the neutrino masses are an order of 
$10^{-6}$ eV for $h \sim O(1)$, 
whereas $m_{\nu} \sim O(1)$ eV for 
$\Lambda \sim 10^{12}$ GeV (which is nothing but the usual intermediate scale
in many beyond standard models).

\vspace{.3in}

{\bf 2.} For the charged lepton sector, we adopt modified 
Fritzsch's form as in Ref.~\cite{nagoya} :
\begin{eqnarray}
M_{l}^{real} = c_{l}~\left( \begin{array}{ccc}
	     0 & d_{l} & 0   \\
	     d_{l} & {2\over 3}~\epsilon_{l} & -{\sqrt{2} \over 3}
	      ~\epsilon_{l} \\
	     0 &  -{\sqrt{2} \over 3}~\epsilon_{l} &  
	     3 + {1 \over 3}~\epsilon_{l} 
	   \end{array} \right),
\end{eqnarray}
in the hierarchical basis where $c_l\simeq m_\tau/3$.  
We assume that $d_{l}\ll \epsilon_{l}\ll 1$. 
Then, one can treat the $d_l$ terms by a small perturbation, and 
solve $d_{l} = 0$ case first.  The corresponding eigenvalues are 
\begin{eqnarray}
\lambda_{0} & = & 0,
\nonumber    \\
\lambda_{\mp} & = & {1\over 2}~\left[ ( 3+ \epsilon_{l} ) \mp 
3~\sqrt{ 1 - { 2 \over 9}~\epsilon_{l} + {1 \over 9}~\epsilon_{l}^{2}}
\right],
\end{eqnarray}
with the eigenvectors 
\begin{eqnarray}
| \lambda_{0} \rangle & = & (1,0,0),
\nonumber  \\
| \lambda_{-} \rangle & = & (0,\cos \theta_{l}, \sin \theta_{l} ),
\\
| \lambda_{+} \rangle & = & (0, -\sin \theta_{l}, \cos \theta_{l} ),
\nonumber  
\end{eqnarray}
where 
\begin{eqnarray}
\sin^{2} \theta_{l} & = & {{2 \epsilon_{l}^{2} } \over 
{ 2 \epsilon_{l}^{2}  + ( 9 + \epsilon_{l} - 
3 \lambda_{-} )^2} },
\nonumber   \\
\cos \theta_{l} & = & { ( 9 + \epsilon_{l} - 3
\lambda_{-} ) \over {\sqrt{2} } \epsilon_{l} }~\sin \theta_{l}.
\end{eqnarray}
At this stage, $m_{e} = 0$ MeV.  To the leading order in $d$, we can identify 
$\lambda_{\mp}$ as $m_{\mu, \tau}$.  Doing so, one gets 
\begin{equation}
\epsilon_{l} = 0.287, ~~~{\rm and}~~~\theta_{l} = 2.66^{\circ}.
\end{equation}
For electron mass (0.511 MeV), a nonzero value of $d_l$ is
needed, $d_{l} = 0.0128~(d_{l}^{2} \approx 1.65 \times 10^{-4})$.
$M_l$ can be approximately diagonalized by a unitary matrix 
\begin{eqnarray}
U_{l} = \left( \begin{array}{ccc}
	1 & {d_{l} \cos \theta_{l} \over \lambda_{-}} & 
	 -{d_{l} \sin \theta \over \lambda_{+}} 
   \\
	-d_{l} \left( {\cos^{2} \theta_{l} \over \lambda_{-}} + 
		      {\sin^{2} \theta_{l} \over \lambda_{+}} \right) & 
	\cos \theta_{l} & -\sin \theta_{l}   \\
	d_{l} \sin \theta_{l} ~\cos \theta_{l} \left( {1 \over \lambda_{+}} - 
   {1 \over \lambda_{-}} \right) & \sin \theta_{l}  & \cos \theta_{l}
	 \end{array} \right).
\end{eqnarray} 
Notice that the angle $\theta_l$ is completely fixed by the measured charged
lepton masses. 
\vspace{.3in}

{\bf 3.} Unlike the charged lepton sector, three neutrinos do not seem to 
have a hierarchy in their masses.  In fact, the solar neutrino puzzle
can be explained through the MSW mechanism if $\Delta m_{solar}^{2} \approx
10^{-5}~{\rm eV}^2$ and $\sin^{2} \theta_{solar} \approx 8 \times 10^{-3}$
(small angle case) , or 
$\sin^{2} \theta_{solar} \approx 0.7$ (large angle case),
and through the just-so vacuum oscillations if $\Delta m_{solar}^{2} \approx
10^{-10}~{\rm eV}^2$.
The atmospheric neutrino problem can be accommodated if 
$\Delta m_{atmos}^{2} \approx 10^{-2}~{\rm eV}^2$ and $\sin^{2} 
\theta_{atmos} \approx 0.5$.   
If light massive neutrinos provide  the hot dark matter of
the universe, one has to require 
\begin{equation}
\sum_{i=1,2,3} | m_{\nu_i} | \sim 6~{\rm eV}.
\end{equation}
All these data indicate that all three neutrinos may be almost 
degenerate in their masses, with $m_{\nu} \sim $~ a few eV, rather than
$m_{\nu_1} \ll m_{\nu_2} \ll m_{\nu_3}$, as usually assumed in the three 
neutrino mixing scenarios \footnote{The recent LSND data \cite{lsnd1}, 
if confirmed,
indicates $\Delta m_{LSND}^{2} \sim 1~{\rm eV}^2$ and $\sin^{2}
\theta_{LSND} \sim 10^{-3}$. Since the conclusions of two different 
analyses \cite{lsnd1,lsnd2} do not agree each other, 
we do not consider the possibility 
alluded by the LSND data \cite{lsnd1} in this work.
See, however, Ref.~\cite{kkk} for a discussion when the LSND data is included.
}. 

Since the flavor democratic neutrino mass matrix leads to a large hierarchy, 
one necessarily has to modify the symmetry relevant to the neutrino mass
matrix.  Here, we propose to consider the permutation symmetry among three
family indices rather than the flavor democracy.  Then, the lowest order
neutrino mass matrix would look like (in the {\it symmetry} basis)
\begin{eqnarray}
{\tilde M}_{\nu}^{(0)} = c_{\nu}~\left( \begin{array}{ccc}
	     1 & r & r   \\
	     r & 1 & r \\
	     r & r & 1
	   \end{array} \right).
\end{eqnarray}
We have assumed that three light neutrinos are  Majorana particles, so that 
the neutrino mass matrix is real symmetric $3 \times 3$ matrix.  
One recovers the flavor democratic case  for $r=1$.   
In the {\it hierarchical} basis, the above form becomes
\begin{eqnarray}
M_{\nu}^{(0)} = c_{\nu}~\left( \begin{array}{ccc}
	     1-r & 0 & 0   \\
	     0 & 1-r & 0 \\
	     0 & 0 & 1+2r    
	   \end{array} \right).
\end{eqnarray}
Either for small $r (\sim 0)$ or for $r \sim -2$,
all the three neutrinos are almost degenerate \footnote{Note that the sign 
of the fermion mass is physically meaningless.}.  
For any $r$, two neutrinos are always degenerate, so that there is only
one $\Delta m_{\nu}^2$ available. 

In order to have two 
$\Delta m^2$ scales from our mass matrices, one has to lift the degeneracy 
for the first two neutrinos further.  
At this stage, there would be two simple 
ways to break this degeneracy :
\begin{eqnarray}
Case\ I :  ~~~~~~~~ M_{\nu}^{(I)} = c_{\nu}~\left( \begin{array}{ccc}
	     1-r & \epsilon_{\nu}  & 0   \\
	     \epsilon_{\nu}  & 1-r & 0 \\
	     0  & 0 & 1+2r
	   \end{array} \right).
\end{eqnarray}
and 
\begin{eqnarray}
Case\ II : ~~~~~~~~ M_{\nu}^{(II)} = c_{\nu}~\left( \begin{array}{ccc}
	     1-r & 0  & 0   \\
	     0  & 1-r & \epsilon_{\nu} \\
	     0 & \epsilon_{\nu} & 1+2r
	   \end{array} \right).
\end{eqnarray}
Other possibilities are equivalent to the above two by changing the labels 
$i=1,2,3$ in the mass eigenstates of neutrinos.  
\vspace{.3in}

{\bf 4.}  Now, we analyze the above two ansatzs for the neutrino mass matrices,
and study the consequences in the neutrino mixing.  
\vspace{.3in}

{\it Case I} : 

In this case, the neutrino mass matrix is diagonalized by 
\begin{eqnarray}
U_{\nu}^{I} = \left( \begin{array}{ccc}
	  -1/ \sqrt{2} & 1/ \sqrt{2} & 0   \\
	   1/ \sqrt{2} & 1/ \sqrt{2} & 0   \\
	   0 & 0 & 1
	 \end{array} \right),
\end{eqnarray} 
with the eigenvalues 
\begin{equation}
m_{\nu_{i}} / c_{\nu} = (1-r \mp \epsilon), (1 + 2r).
\end{equation}
Note that we have shown only one possibility for labeling three neutrino
mass eigenstates.  Furthermore, Eq. (12) is independent of neutrino mass 
eigenvalues, or equivalently, on $c_{\nu}, \epsilon_{\nu}$ and $r$. 
This is because the mass matrix $M_{\nu}^{(I)}$ given in Eq. (10) 
still has a residual
$S_2$ symmetry acting upon the first and the second family indices.  

Combining with the $U_{l}$  given in (6), 
one gets the neutrino mixing matrix,
$V^{I} \equiv U_{l}^{\dagger} ~U_{\nu}^{I}$.  First of all, the mixing matrix
is independent of neutrino masses, although it depends on the charged
lepton masses.  One can solve for $c_{\nu}, r$ and $\epsilon_{\nu}$
by requiring three conditions, $\Delta m_{solar}^2 \simeq 10^{-10}~
{\rm eV}^2$, $\Delta m_{atmos}^2 \simeq 0.72 \times 10^{-2} ~{\rm eV}^2$ 
and Eq. (7) \footnote{In this work, we solve the solar neutrino problem 
in terms of vacuum oscillations. The results would remain the same even 
if we invoke the MSW mechanism.}. Then, we check if the 
solution satisfies the constraint from the neutrino-less double $\beta-$decay,
and other data from neutrino oscillation experiments.   

We find that there are 
three sets of parameters, $(c_{\nu}, r, \epsilon_{\nu})$ :
\begin{eqnarray}
& & (1.0, \pm 0.667, \mp 2.997)
\nonumber   \\
(r, c_{\nu}, \epsilon_{\nu}) & = & (1.001, \pm 0.666, \mp 3.004)
\\
& & (0.999, \pm 0.667, \mp 2.999).
\nonumber  
\end{eqnarray}
Since we are considering Majorana neutrinos, 
there is an additional constraint from
non-observation of neutrino-less double $\beta-$decays \cite{doublebeta} :
\begin{equation}
\langle m_{\nu_e} \rangle \equiv | \sum_{i=1}^{3} ~ V_{ei}^{2} m_{i} |
< 0.7~{\rm eV}.
\end{equation}
This condition is not easy to satisfy  in typical models with massive
neutrinos when one tries to solve the hot dark matter problem in terms
of light neutrinos with masses in a few eV range, as discussed in
Ref.~\cite{langacker}.

The values for $\langle m_{\nu_e} \rangle$ corresponding to
Eq. (14) are
\begin{equation}
\langle m_{\nu_e} \rangle = 0.2771~{\rm eV}, ~~ 0.2763~{\rm eV}
 ~~{\rm and}~~0.2781~{\rm eV},
\end{equation}
respectively. All of these solutions  are 
well below the current upper limit given in Eq. (15). 
The expected atmospheric neutrino data $R$'s  for different
$L/E$ from the atmospheric are given in Table~1 along with the 
current data from KAMIOKANDE, IMB, FREJUS, NUSEX and SOUDAN. From
Table ~1, we find that the lepton mass matrix ansatz
Eqs. (11) and (14) reproduce all the known data on the neutrino oscillation 
experiments.  

Further test of our ansatz is provided with the long baseline experiments
searching for $\nu_{\mu} \rightarrow \nu_{\tau}$ oscillation in the range 
of  $\Delta m_{\nu}^2 \simeq 10^{-2}~{\rm eV}^2$.  Our prediction is that 
\begin{equation}
P ( \nu_{\mu} \rightarrow \nu_{\tau} ) \sim 7.6 \times 10^{-3},
\end{equation}
with $\Delta m^2 = 0.72 \times 10^{-2}~~{\rm eV}^2$. 
This is still well below the current upper limit as well as the planned 
search for the $\nu_{\mu} \rightarrow \nu_{\tau}$ oscillations at 
CHORUS, NOMAD, ICARUS, etc.
\vspace{.3in}

{\it Case II} :

In this case, the neutrino mass matrix is diagonalized by
\begin{eqnarray}
U_{\nu}^{II} = \left( \begin{array}{ccc}
	  1 & 0 & 0   \\
	  0 & \cos \theta_{\nu} & -\sin \theta_{\nu}   \\
	  0 & \sin \theta_{\nu}  & \cos \theta_{\nu}    
	 \end{array} \right),
\end{eqnarray} 
with the mass eigenvalues being
\begin{equation}
m_{i}/c_{\nu} = 1-r,~~~{1 \over 2}~\left[ ( 2 + r ) \pm \sqrt{ 9 r^{2} + 
4 \epsilon_{\nu}^{2}} \right].
\end{equation}
The mixing angle $\theta_{\nu}$ is determined by
\begin{equation}
\sin^{2} \theta_{\nu} = {{2 \epsilon_{\nu}^{2}} \over 
(9 r^{2} + 4 \epsilon_{\nu}^{2} ) - 3 r \sqrt{ 9 r^{2} + 4 
\epsilon_{\nu}^{2} } },
\end{equation}
with
\begin{equation}
\cos \theta_{\nu} = \left( {{ \sqrt{  9 r^{2} + 4   \epsilon_{\nu}^{2} } 
- 3 r } \over 2 \epsilon_\nu} \right)~\sin \theta_{\nu}.
\end{equation}
Unlike the {\it Case I}, we now have the neutrino mixing matrix which does 
depend on the neutrino masses in a nontrivial way.  

For a given set of 
$m_{i}^2$, one can get $\theta_{\nu}$.  Scanning the parameter space 
in this case as before, we find no solution. 
Typically, $| V_{1e}| \sim 1$ and $ | V_{2e}| , | V_{3e} | \ll 1$, we have 
too large $\langle m_{\nu_e} \rangle \simeq 2~ $eV to be compared to
(15).  Thus, the neutrino mass ansatz Eq. (11) along with 
the charged lepton mass ansatz Eq. (1) 
 does not fit the neutrino data.
\vspace{.3in}

{\bf 5.} In conclusion, we investigated 
the lepton mass matrices with the minimal 
number of parameters, three in each of the charged lepton 
and Majorana neutrino mass matrices ($M_{l}$ and $M_{\nu}$), 
with a permutation symmetry among three 
generations ($S_3$) and its suitable breaking into $S_1$ and $S_2$, 
respectively.
We find the ansatz (1) and (10) lead to a lepton mixing matrix which is
consistent with the current data on various types of neutrino
oscillation experiments. Three light Majorana neutrinos can serve as  
the hot dark matter, with $\Sigma | m_{\nu_i} | \sim 6~$ eV.  The resulting
$\nu_{\tau} \leftrightarrow \nu_{\mu}$ probability is in the range of
0.008 with $\Delta m^2 \simeq 0.7 \times 10^{-2}~{\rm eV}^2$,  which still 
lies beyond the scope of the planned CHORUS and other experiments searching 
for $\nu_{\tau} \leftrightarrow \nu_{\mu}$ oscillation. Furthermore, three
neutrinos being  almost degenerate, we expect that the lepton family number
breaking will be very small.  

\acknowledgements

Two of us (JEK,PK) thank the Brown High Energy Theory Group for 
the hospitality during the visit.
This work is supported in part by the Korea Science and Engineering 
Foundation through Center for Theoretical Physics at Seoul National
University (SKK, JEK, PK), SNU-Brown Exchange Program (KK, JEK, PK),
Korea-Japan Exchange Program (JEK), the
Ministry of Education through the Basic Science Research Institute,
Contract No. BSRI-96-2418 (JEK,PK),
and also the US DOE Contract DE-FG-02-91ER40688 - Task A (KK).


%
%
%
%
%
\begin{table}
\caption{The atmospheric neutrino data $R$ for various $L/E$ along with
our predictions for $\Delta m_{21}^{2} = 0.72 \times 10^{-2}~{\rm eV}^2$, 
$\Delta m_{32}^{2} = 
 \times 10^{-10}~{\rm eV}^2$
We show the $r = (\mu/e)_{\rm incident}$ 
values for each data point also.}
\label{table1}
\begin{tabular}{ccccc}
Experiments & $r$ & $L/E$ (km/GeV) & Measured & Prediction 
\\   \tableline
KAMIOKA \cite{kamioka} & $4.5/1$ & 5  & $1.27^{+0.61}_{-0.38}$ & 
0.99
\\
(Multi-GeV)            & $3.2/1$ & 10 & $0.63^{+0.21}_{-0.16}$ & 
0.97
\\
		       & $2.2/1$ & 100& $0.51^{+0.15}_{-0.12}$ & 
0.41
\\
		       & $3.2/1$ & 1000 & $0.46^{+0.18}_{-0.12}$ & 
0.31
\\
		       & $4.5/1$ & 2000 & $0.28^{+0.10}_{-0.07}$ & 
0.22
\\
KAMIOKA \cite{kamioka} & $2.1/1$ & 80   & $0.59 \pm 0.10$ & 
0.50
\\
(Sub-GeV)              & $2.1/1$ & 12800 & $0.62 \pm 0.10$  &
0.48
\\
IMB \cite{imb}         & $2.1/1$ & 1000 &  $0.54 \pm 0.13$ & 
0.47
\\
FREJUS \cite{frejus}   & $2.1/1$ & 500  &  $0.87 \pm 0.18$ &   
0.47
\\
NUSEX \cite{nusex}     & $2.1/1$ & 500  &  $0.99 \pm 0.32$ & 
0.47
\\
SOUDAN \cite{soudan}   & $2.1/1$ & 1000 & $0.69 \pm 0.21$  & 
0.47
\end{tabular}
\end{table}

\end{document}